\documentclass[useAMS,referee,usenatbib,table]{biom} 

\usepackage[table,dvipsnames]{xcolor}
\usepackage[T1]{fontenc}
\usepackage[utf8]{inputenc}
\usepackage{amssymb,amsmath}
\usepackage{bbm}
\usepackage{comment}
\usepackage[pdfusetitle]{hyperref}
\usepackage[all]{xy}
\usepackage{graphicx}
\usepackage{natbib}
\usepackage{multirow} 
\usepackage{pdflscape}
\usepackage{setspace}
\usepackage{algorithm}
\usepackage{algpseudocode}
\usepackage{tabularx}
\usepackage{booktabs}
\usepackage{ragged2e}
\usepackage{blindtext}
\newtheorem{thm}{Theorem}

\usepackage[skip=2pt,font=footnotesize]{caption}

\newcommand{\indep}{\perp \!\!\! \perp}
\newcommand{\norm}[1]{\left\lVert#1\right\rVert}

\title[]{A Doubly Robust Instrumental Variable Approach for Estimating Average Treatment Effects in Time-to-Event Data with Unmeasured Confounding: Application to Real-World Data on ICU Patients with Septic Shock}
\author{Runjia Li$^{1}$, Victor B. Talisa$^{2}$, and Chung-Chou H. Chang$^{1,3}$ \\
$^{1}$Department of Biostatistics and Health Data Science, University of Pittsburgh, Pittsburgh, Pennsylvania \\
$^{2}$Department of Critical Care, University of Pittsburgh, Pittsburgh, Pennsylvania\\
$^{3}$Department of Medicine, University of Pittsburgh, Pittsburgh, Pennsylvania}

\date{}
\begin{document}
\label{firstpage}

\begin{abstract}
Motivated by conflicting conclusions regarding hydrocortisone's treatment effect on ICU patients with vasopressor-dependent septic shock, we developed a novel instrumental variable (IV) estimator to assess the average treatment effect (ATE) in time-to-event data. In real-world data, IV methods are widely used for estimating causal treatment effects in the presence of unmeasured confounding, but existing approaches for time-to-event outcomes are often constrained by strong parametric assumptions and lack desired statistical properties. Based on our derived the efficient influence function (EIF), the proposed estimator possesses double robustness and achieves asymptotic efficiency. It is also flexible to accommodate machine learning models for outcome, treatment, instrument, and censoring for handling complex real-world data. Through extensive simulations, we demonstrate its double robustness, asymptotic normality, and ideal performance in complex data settings. Using electronic health records (EHR) from ICU patients, we identified physician preferences for prescribing hydrocortisone as the IV to evaluate the treatment effect of hydrocortisone on mortality. Our analysis shows no significant benefit or harm of hydrocortisone on these patients. Applying the proposed doubly robust IV method provides reliable estimates in the presence of unmeasured confounders and offers clinicians with valuable evidence-based guidance for prescribing decisions.\\
\end{abstract}

\maketitle
\clearpage
\pagenumbering{arabic} 

\section{Introduction}\label{c3s:intro}
Accurately estimating causal treatment effects is critical in biopharmaceutical research. However, the unconfoundedness assumption is often violated in real-world data (RWD) and randomized controlled trials with non-compliance, leading to biased estimates. For example, conflicting conclusions about the effect of hydrocortisone on patients with septic shock are likely due to unmeasured confounders, such as medical history and other hard-to-measure factors. This motivates developing robust methods to address unmeasured confounding. Instrumental variable (IV) methods offer a solution using instruments that influence treatment assignment without directly affecting the outcome, except through the treatment. However, most existing IV methods for time-to-event data rely on structural Cox models, requiring correct model specification and proportional hazards assumptions \citep{mackenzie2014using,martinez2019adjusting,sorensen2019causal}. Moreover, targeting hazard ratios also can yield inconsistent estimates due to the noncollapsibility issue \citep{wang2023instrumental}. 

Alternative methods incorporating additive hazards models into two-stage regression frameworks \citep{li2015instrumental, tchetgen2015instrumental, ying2019two, martinussen2017instrumental} partially address these issues but impose other restrictive assumptions, limiting their applicability in complex RWD. To overcome this, nonparametric estimators for local average treatment effects (LATE) have been developed, focusing on the subpopulation complying with treatment assignment \citep{frandsen2015treatment,yu2015semiparametric,lee2023doubly}. However, LATE are often not of primary interest, as they only focus on a subset of the population.

In this paper, we propose a novel doubly robust estimator for the average treatment effects (ATE) in time-to-event data using an IV framework to address unmeasured confounding. This paper makes the following contributions: (1) we define the ATE as the difference in cumulative incidence functions (CIFs), extendable to settings with competing risks, and identify it from observed data using an instrument under necessary assumptions; (2) We derive an estimator based on the efficient influence function (EIF) that achieves double robustness, asymptotic efficiency, and valid inference; and (3) The proposed method accommodates flexible integration of statistical and machine learning models for CIF, censoring, treatment assignment, and instrument prevalence to handle complex data. 
By addressing unmeasured confounding with the proposed flexible and robust approach, our work fills a critical gap in causal inference for time-to-event data. It provides a valuable tool for estimating treatment effects in real-world biopharmaceutical applications, advancing both methodological and clinical decision-making frontiers.

This article is organized as follows. In Section \ref{c3s:set}, we introduce the notation and define the ATE. Section \ref{c3s:identification} outlines the assumptions for identifiability. In Section \ref{c3s:method}, we demonstrate the proposed one-step estimator and its asymptotic properties. Section \ref{c3s:sim} presents simulations to evaluate the performance of our estimator and compare it with alternative methods. In Section \ref{c3s:app}, we describe a real-world application. Finally, we conclude in Section \ref{c3s:dis}.

\section{Setting and Notation}\label{c3s:set}
\subsection{Time-to-event outcomes}\label{c3ss:notation}
\begin{figure}
\centerline{\includegraphics[width=0.45\linewidth]{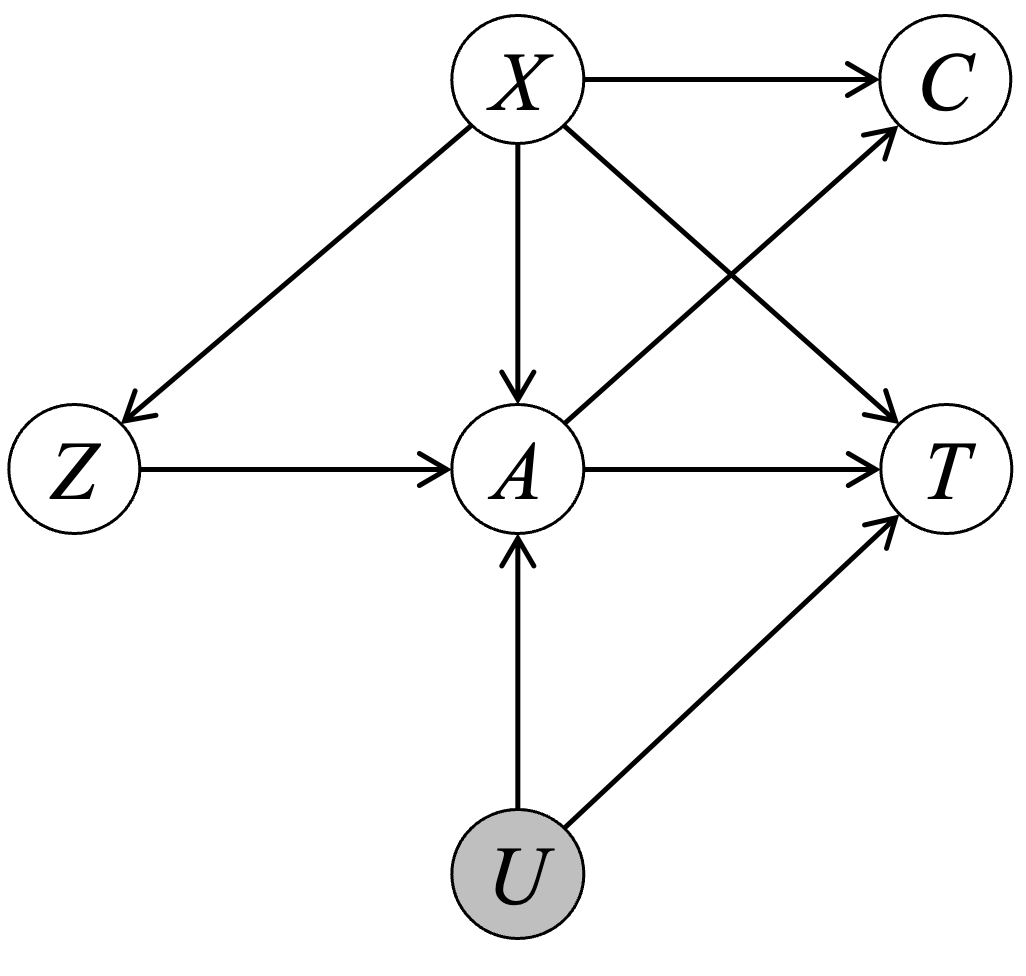}}
\caption[The causal diagram of time-to-event data with unmeasured confounding.]{The relationship between treatment $A$, observed confounders $X$, instrument variable $Z$, unmeasured confounders $U$, event time $T$, and noninformative right-censoring time $C$.}
\label{c3f:dag}
\end{figure}

We consider an observational study with time-to-event outcomes with two treatment arms. The observed data is denoted as $O_i=\left(\Tilde{T}_i, {\Delta_i}, A_i, X_i,Z_i\right)_{i=1}^n \sim^{i.i.d.} P$ for subject $i=1, \ldots, n$, where $\tilde T=T\wedge C$ is the observed event time, $T$ is the failure time, and $C$ is the right-censoring time. $\Delta=\mathbbm{1}(T\leq C)$ is the event indicator. $X\in\mathcal{X}\subseteq\mathbb R^p$ represents $p$-dimensional  baseline covariates, $A\in \{0,1\}$ is the binary treatment assignment, $U$ is the baseline unmeasured confounders, and $Z\in \{0,1\}$ is the binary instrumental variable. $P$ denotes the unknown true distribution of $O$, from which the observed sample is drawn. The relationships among $A$, $X$, $Z$, $U$, $T$, and $C$ can be depicted by a directed acyclic graph (DAG) in Figure \ref{c3f:dag}.

Let $R(t)=\mathbbm{1}\left(T \leq t\right)$ be the failure indicator by time $t$, and the cumulative incidence function (CIF), also called the cumulative failure probability, is $F(t)=\operatorname{Pr}(T\leq t)=E\{R(t)\}$. 

\subsection{Causal estimand} \label{c3ss:causalest}
To define and identify our target causal estimand, we adopt the potential outcome framework \citep{neyman1923application,rubin1974estimating}. Let $T^a$ and $C^a$ be the potential event time and censoring time, respectively, if the subject had received treatment $A=a$ $(a=0,1)$. Then, the potential failure indicator at time $t$ is $R^a(t)=I(T^a\leq t)$, and the potential CIF at time $t$ is $F^a(t)=\operatorname{Pr}(T^a\leq t)=E\{R^a(t)\}$, reflecting scenarios that might not actually occur. 

Our causal estimand, the ATE, is defined as the difference in potential CIFs:
$$
\Psi_t(P)=F^1(t)-F^0(t)=E\{R^1(t)-R^0(t)\}.
$$
For simplicity, we denote $R^a(t)$ as $R^a$.

\section{Identification of Average Treatment Effect}\label{c3s:identification}
To identify the ATE, we employ an instrumental variable $Z$ that satisfies a set of assumptions. Let $A^z$ be the potential treatment that the subject would receive if the instrument had been set to $Z=z$ $(z=0,1)$, and $R^{a,z}$ denote the potential outcome when the instrument is set to $Z=z$ and the subject is assigned to $A=a$. We assume that $A^z$ and $R^{a,z}$ are not affected by the instrument or treatment of other subjects, and only one version of the treatment and instrument exists. The following assumptions are required identifying the ATE:
\begin{itemize} 
    \item []{Assumption A1 (Consistency).} The potential outcome $R^{a,z}$, is equal to the observed outcome when treatment $A = a$ and instrument $Z=z$; and the potential treatment assignment $A^z$, is equal to the observed treatment when the instrument $Z=z$.
    \item []{Assumption A2 (Relevance).} The instrument relates to the treatment: $cov(Z,A|X)\neq 0 $.
    \item []{Assumption A3 (Positivity).} Positive probabilities of each instrument and treatment value, and getting censored, i.e., 
    $\operatorname{Pr}(Z=z|X)>0, \operatorname{Pr}(A=a|X,Z)>0, $ and $\operatorname{Pr}(C>t|X,A)>0$. 
    \item []Assumption A4 (Noninformative Censoring). The censoring time $C$ is independent of the failure time $T$ conditional on the covariates $X$ and treatment $A$. $T\indep C|(X,A)$.
    \item []Assumption A5 (Exclusion Restriction). The instrument $Z$ affects the outcome $R$ only through its effect on the treatment assignment $A$, i.e., $R^{z,a}=R^a$ for all $z$.
\end{itemize}

The following assumptions, A6 and A7, enable us to identify the ATE. They closely mirror conditions A5a and A5b in \cite{qiu2021optimal}, which state the exchangeability given all confounders and the absence of unmeasured treatment-outcome effect modifiers, respectively.
\begin{itemize}
    \item []Assumption A6 (Exchangeability). $R^a\indep (Z, A)|X,U$. 
    \item []Assumption A7. At least one set of the following is satisfied:
    \begin{enumerate}
        \item Both assumptions hold:
        \begin{enumerate} 
            \item (Independent instrument) $Z\indep U|X$ almost surely (a.s.)
            \item (No unmeasured instrument-treatment effect modification) \\$E(A|Z=1,X,U)-E(A|Z=0,X,U)=E(A|Z=1,X)-E(A|Z=0,X)$ a.s.
         \end{enumerate}
         \item Both assumptions hold:
         \begin{enumerate}
            \item (Uncorrelated instrument) $cov(R^0,Z|X)=0 $ a.s.
            \item (No unmeasured treatment-outcome effect modification) $E(R^1-R^0|X,U)=E(R^1-R^0|X) $ a.s.
         \end{enumerate}
    \end{enumerate}
\end{itemize}
\begin{thm}\label{c3thm:iden}
Under Assumptions A1-A6, and if either part of A7 holds, the ATE can be identified via the average of the conditional Wald estimand \citep{wald1940fitting}:
 $$\Psi_t(P)=E\left(R^1-R^0\right)=E_X\left\{\frac{E(R|Z=1,X)-E(R|Z=0,X)}{E(A|Z=1,X)-E(A|Z=0,X)}\right\}.$$
\end{thm}
The proof for Theorem \ref{c3thm:iden} is provided in Supplementary Materials.
\section{Proposed Estimator}\label{c3s:method}
After establishing identification of the ATE in Section \ref{c3s:identification}, we proceed to construct nonparametric estimators. For ease of notation, let $\Psi^R_{Z=z,X}=E(R|Z=z,X)$ represent the CIF conditional on the instrument $Z=z$ and covariates $X$, and let $\Psi^A_{Z=z,X}=E(A|Z=z,X)$ denote the probability of being assigned treatment given $Z=z$ and covariates $X$.
\subsection{A naive plug-in estimator and plug-in bias}
\setlength{\emergencystretch}{0.5\textwidth}\par
As the ATE is identified by the expectation of the conditional Wald estimand $E_X\left\{(\Psi^R_{Z=1,X}-\Psi^R_{Z=0,X})/(\Psi^A_{Z=1,X}-\Psi^A_{Z=0,X})\right\}$, one might consider a naive plug-in estimator by plugging in the estimated CIF ($\hat\Psi^R_{Z=z,X}$), and the estimated treatment probabilities ($\hat \Psi^A_{Z=z,X}$). The empirical mean of these estimations serves as the plug-in estimator: 
\begin{equation}\label{c3eq:plugin}
\Psi_{\text{pi}}(\hat P_n)=\mathbb{P}_{n}\left(\frac{\hat \Psi^R_{Z=1,X}-\hat \Psi^R_{Z=0,X}}{\hat \Psi^A_{Z=1,X}-\hat \Psi ^A_{Z=0,X}}\right),
\end{equation}
where $\mathbb{P}_{n}$ denotes the empirical average $\mathbb{P}_{n}(\cdot)={n^{-1}}\cdot\sum_{i=1}^{n}(\cdot)$. However, this naive plug-in estimator introduces a first-order bias, as shown by the following equation derived from the von Mises expansion \citep{hines2022demystifying, kennedy2022semiparametric}:
\begin{equation}\label{c3eq:pibias}
\Psi_{\text{pi}}(\hat P_n)-\Psi(P)=-E\{D(\hat P_n)\}+R(\hat P_n,P),
\end{equation}
where $D(P)$ is the influence function, with $E\{D( P)\}=0$ and $E\{D(P)^2\}<\infty$. $R(\hat P_n,P)$ is a second-order remainder term of the expansion, and will be further discussed in Section \ref{c3ss:est}. To correct the first-order bias term $-E\{D(\hat P_n)\}$, we subtract the empirical average $-\mathbbm{P}_n\{D(\hat P_n)\}$ from the plug-in estimator, as detailed in the next sections.
\vspace{-0.5em}
\subsection{Efficient influence function (EIF)}\label{c3ss:eif}
To correct the plug-in bias and achieves asymptotic efficiency, we need to construct our estimator based on the EIF of $\Psi(P)$. Let $D(P)$ be the EIF of the estimator. To introduce the EIF, we need the following nuisance functions: (1) the CIF $F(t|Z,X)$, (2) the probability of treatment assignment $\pi(a|Z,X)=\operatorname{Pr}(A=a|Z,X)$, (3) the censoring probability $G(s|Z,X)=\operatorname{Pr}(C>s|Z,X)$, and (4) prevalence of the instrument $\gamma(z|X)=\operatorname{Pr}(Z=z|X)$.
\vspace{-0.5em}
\begin{thm}\label{c3thm:eif} (Efficient Influence Function of ATE) \normalfont Under Assumptions A1-A6 and either A7(1) or A7(2), the EIF of the ATE $\Psi(P)$ is given by  
\begin{equation}\label{c3e:eif}
\begin{aligned}
D(P)=\frac{D^R}{\Psi_X^A}-\Psi_X \cdot\frac{D^A}{\Psi_X^A}+\Psi_X-\Psi,
\end{aligned}
\end{equation}
where $\Psi_X^A=\Psi^A_{Z=1,X}- \Psi ^A_{Z=0,X}$, $\Psi_X^R= \Psi^R_{Z=1,X}-\Psi ^R_{Z=0,X}$, $\Psi_X=\Psi_X^R/\Psi_X^A$, 
$$
\begin{aligned}
D^R&=\int_0^t\frac{(2Z-1)}{\gamma(Z|X)}\frac{\left\{1-F(t|Z,X)\right\}}{\left\{1-F(s|Z,X)\right\}G(s^-|Z,X)}M(ds|Z,X), \text{ and} \\
D^A&= \frac{2Z-1}{\gamma(Z|X)}\left\{I(A=1)-\pi(1|Z,X)\right\},
\end{aligned}
$$
and $M(ds|z,x)=N(ds) -\mathbbm{1}(\tilde{T}\geq s)\Lambda(ds|z,x)$, and $N(s)=\mathbbm{1}(\tilde{T}\leq s,\Delta=1)$.
\end{thm}
The proof of Theorem \ref{c3thm:eif} is provided in Supplementary Materials. 

\subsection{Doubly robust estimator and its asymptotic properties}\label{c3ss:est}
With the form of EIF, we obtain the one-step estimator by substracting the empirical average of EIF from the plug-in estimator:     
\begin{equation}\label{os}
    \Psi_{os}(\hat P_n)=\mathbb{P}_n\left(\hat\Psi_X+\frac{\hat D^R}{\hat \Psi_X^A}-\hat\Psi_X\frac{\hat D^A}{\hat \Psi_X^A}\right).
\end{equation}

To illustrate the asymptotic properties of the estimator, consider conditions C1-C2:
\begin{itemize}
    \item []Condition C1. All estimators of the nuisances belong to the Donsker class, which includes functions that are not overly complex, such as traditional parametric functions, indicator functions, and certain smooth machine learning estimators like LASSO and generalized additive models (GAM) \citep{van2011targeted,balzer2023invited}. 
    \item []Condition C2. We assume $0<\hat F(t|Z,X)<1$, $\hat G(t^-|A,X)>0$, and $\hat \gamma(z|X)>0$ to ensure the estimator is well-defined with non-zero denominators.
\end{itemize}
Condition C1 restricts model complexity to prevent overfitting. However, some machine learning methods, such as random forest and deep neural networks, do not belong to the Donsker class. To allow more flexibility in modeling nuisance functions and capture complex relationships in the data, we also consider sample splitting as an alternative approach to C1 for avoiding overfitting \citep{zheng2010asymptotic,chernozhukov_doubledebiased_2018}. The specific Donsker class models and sample splitting methods will be discussed in next section.
\begin{thm}\label{c3thm:asy}
Under Conditions C1 and C2, as well as Assumptions A1-A7, the asymptotic properties of the proposed estimator (\ref{os}) $\Psi_{os}(\hat P_n)$ are described by the following equation:
\begin{equation}\label{c3eq:remainder}
    \begin{aligned}
     &\Psi_{os}(\hat P_n)-\Psi(P) \\
      &= O_p(\sum_{z\in\{0,1\}}\left\{\norm{\int_0^t\gamma(z|X)G(s^-|z,X)-\hat \gamma(z|X)\hat G(s^-|z,X)ds}\cdot\norm{F(t|z,X)-\hat F(t|z,X)}\right.\\&\quad \quad\left.+\norm{\gamma(z|X)-\hat \gamma(z|X)}\cdot\norm{\pi(1|z,X)-\hat\pi(1|z,X)}\right\}+\norm{\hat \Psi_X-\Psi_X}\cdot\norm{\Psi_X^A-\hat\Psi_X^A})\\&\quad\quad+
      o_p(n^{-1/2}).
    \end{aligned}
\end{equation}
\end{thm}
The proof of Theorem \ref{c3thm:asy} is in Supplementary Materials. This theorem shows the structure of the second-order remainder, where the double robustness of the proposed one-step estimator stem from. It encompass products of the error terms of nuisances, implying that it converges faster than each nuisance function estimator individually, and the asymptotic efficiency can be achieved. Specifically, the proposed one-step estimator is $o_p(n^{-1/2})$ when the all three product of error terms of the nuisance functions in (\ref{c3eq:remainder}) are $o_p(n^{-1/2})$. It provides protection against misspecification in any nuisance model, as long as other components are $o_p(n^{-1/2})$.  


When the proposed one-step estimator is $o_p(n^{-1/2})$ under these mild conditions, by the Central Limit Theorem, $\Psi_{os}(\hat P_n)$ is asymptotic normal:
\begin{equation*}
\sqrt{n}\left\{\Psi_{os}(\hat P_n)-\Psi(P)\right\}\xrightarrow{\mathcal{D}} N\left(0,E_{P}\{D(P)^2\}\right).
\end{equation*}

The asymptotic variance is estimated by $\hat\sigma ^2=\mathbb{P}_n\{D(\hat P_n)\}^2$, and the Wald confidence interval is $\Psi_{os} (\hat P_n)\pm z_{1-\alpha/2}\sqrt{\mathbb{P}_n\{D(\hat P_n)\}^2}$, where $z_{1-\alpha/2}$ is the ${1-\alpha/2}$ percentile of the standard normal distribution.

\subsection{Estimation of nuisances}\label{c3ss:nui}
Our proposed one-step estimator allows for the use of various statistical or machine learning models for the CIF, treatment mechanisms, censoring mechanisms, and instrument prevalence, provided they meet the Donsker condition in Condition C1. To ensure nuisance functions fall into the Donsker class, we consider the following models: for binary treatment and binary instrumental variable, we use generalized linear models (GLMs) and general additive models (GAMs); for predicting CIF and/or censoring probability, we can use the Cox proportional hazards model and/or penalized Cox model \citep{goeman2010l1}. 

As an alternative to the Donsker condition in C1, we use sample splitting to incorporate complex machine learning models and capture complicated relationships in the data. Specifically, the data is randomly partitioned into $K$ equal-sized subsets. For each subset $k$ from $1$ to $K$, we estimate the nuisance functions and EIF using the remaining data, and then calculate the ATE for that subset, denoted as $\hat\Psi_k$. After iterating through all subsets, the final ATE is obtained as $\hat\Psi _k=\frac{1}{K}\sum_{k=1}^{K}\hat\Psi_k$. This approach allows for more flexible modeling of nuisance functions. For binary treatment and instrumental variable, we can use random forest or super learner algorithm; for predicting CIF and censoring probability, we consider the super learner for survival data, random survival forests, and deep survival neural network.

\section{Simulation Studies}\label{c3s:sim}
To evaluate the performance of our proposed one-step estimator, we conducted simulation studies across various settings. The data generating schema include true models for instrument prevalence, treatment probability, event time, and censoring time. Four baseline measured confounders were generated, with $X_1, X_2\sim ^{i.i.d}Unif(-1,1)$ and $X_3, X_4\sim^{i.i.d.}N(0,1)$, and the unmeasured confounder $U\sim Unif(0,1)$. Treatment, event time, censoring time, and instrumental variable were generated from different true models in three simulation sets.
For each setting, we calculated bias and root-mean-square error (RMSE) based on $B=500$ Monte-Carlo repetitions at the 30th, 40th, 50th and 60th percentile event times: $\text{Bias}(t)=\Psi_t(\hat{P}_n)-\Psi_t(P) \text{ and RMSE }(t)=\sqrt{\frac{1}{B}\sum^B_{b=1}\{\Psi_t(\hat{P}_n)-\Psi_t(P)\}^2}$, where $\Psi_t(\hat{P}_n)$ and $\Psi_t(P) $ represent the estimated and true ATE at time $t$, respectively, and $b=1,2,...,B$. 

We also calculated 95\% confidence intervals (CIs) and the proportion of true ATE covered by the 95\% CI to assess asymptotic normality. The CIs for the G-formula and plug-in estimators were based on sampling standard errors, while CIs for the AIPTW and the one-step estimators used the asymptotic variance estimated from the EIFs.

\subsection{Simulation set 1: Weak vs. strong confounding effects}\label{c3ss:sim1}
To assess the performance of the proposed estimators under weak and strong unmeasured confounding, we varied coefficients of $U$ in the true models for treatment and event time. 
The binary instrument $Z$ was generated from $Ber([1+\exp\{-(0.4X_1-0.3X_2-0.1X_3-0.2X_4)\}]^{-1})$. For the data with weak unmeasured confounding effects, treatment $A$ was from $Ber(\left[1+0.1\exp\left\{-5(-X_1-2X_2+0.5X_3-2X_4)\right\}\right]^{-1}+0.855(Z-0.5)+0.045(U-0.5)+0.45)$, with event and censoring times from exponential distributions with means $0.1\exp(-0.2X_1+0.4X_2-0.6X_3+0.4X_4-0.5A+0.6U)$ and $0.05\exp(-0.5X_1+1.5X_2+0.1X_3-0.5X_4-1.5A)$, respectively. For strong unmeasured confounding, treatment $A$ was from $Ber(\left[1+0.1\exp\left\{-5(-X_1-2X_2+0.5X_3-2X_4)\right\}\right]^{-1}+0.63(Z-0.5)+0.07(U-0.5)+0.45)$, with event and censoring times from exponential distributions with means $0.1\exp(-0.1X_1+0.3X_2+0.1X_3-0.1X_4-2A+4U)$ and $0.1\exp(-0.5X_1+1.5X_2+0.1X_3-0.5X_4-0.5A)$.  

We evaluated and compared the bias, RMSE, and coverage probability of the one-step estimator (\ref{os}) and the plug-in estimator (\ref{c3eq:plugin}) with the G-formula estimator and the augmented inverse probability of treatment weighted (AIPTW) estimator for right-censored data \citep{ozenne_estimation_2020}. The G-formula calculates the ATE as the difference between the CIFs in treated and untreated groups, predicted by Cox models that include measured confounders and treatment as covariates. The AIPTW estimator for right-censored data developed by \cite {ozenne_estimation_2020} is a doubly robust estimator derived from semiparametric theory-based estimating equations. Both the G-formula and AIPTW estimators do not account for unmeasured confounding effect and were implemented using the R package \texttt{riskRegression}.

\begin{figure} 
\centerline{\includegraphics[width=0.75\textwidth]{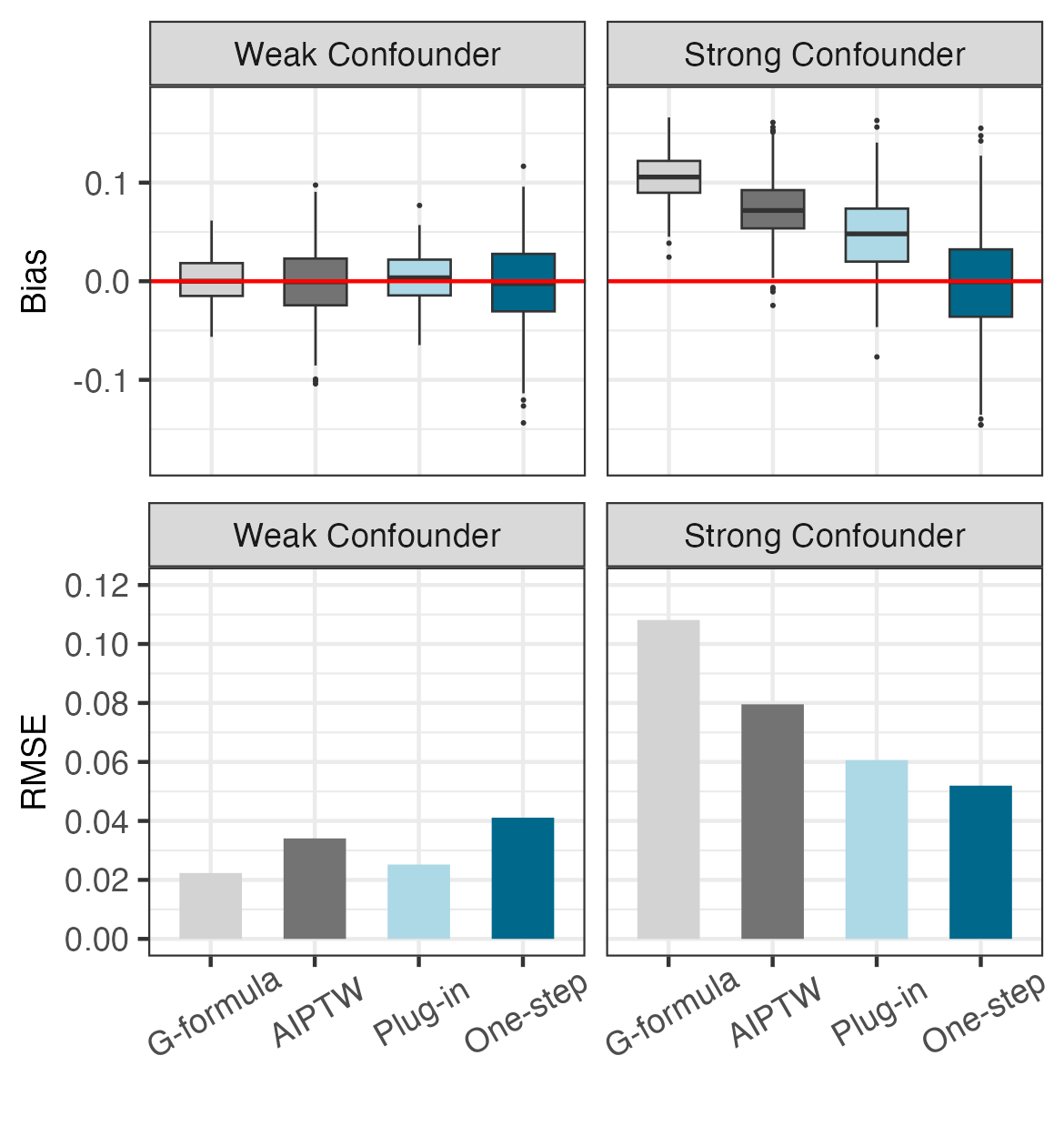}}
\caption[Performance of estimating ATE with weak and strong confounding effects.]{Performance of estimators for estimating ATE at the 50th percentile follow-up time, under both weak and strong confounding effects. The G-formula and AIPTW estimators calculate the ATE without accounting for the unmeasured confounder, while the plug-in estimator and the proposed one-step estimator adjust for the unmeasured confounding effect by using an instrumental variable. All estimators use Cox model to estimate the CIF. For each estimator, censoring probability was estimated using a Cox model, while treatment probability and instrument probability were estimated using logistic regression when applicable. Bias and RMSE were calculated based on 500 simulated datasets, each with a sample size of 1000. The red line represents zero bias.}
\label{c3f:AdjustU}
\end{figure} 

Figure \ref{c3f:AdjustU} the bias and RMSE of the estimators at the 50th follow-up point. The left panel represents weak unmeasured confounding, and the right panel represents strong unmeasured confounding. With weak confounding, all estimators provide unbiased estimates with low RMSE. The plug-in and G-formula estimators perform similarly, while the one-step and AIPTW estimators show slightly higher variance, as expected for doubly robust methods. 

In the right panel with strong unmeasured confounding, the one-step estimator outperforms all other estimators. The G-formula and AIPTW estimators exhibit significant bias, as AIPTW’s double robustness cannot account for misspecification in all models. The plug-in estimator, employing the instrumental variable, shows less bias than the G-formula and AIPTW, but still suffers from bias, especially at earlier time points due to challenges in estimating nuisance parameters with fewer events, as shown in Supplementary Materials. In contrast, the one-step estimator provides unbiased estimates with low RMSE, effectively accounting for unmeasured confounding, whether weak or strong.

\begin{table}
\caption{Coverage probability of ATE estimations at multiple follow-up time points, under weak and strong unmeasured confounding effects.}\label{c3tb:adjustUcov}
\begin{center}
\resizebox{0.75\textwidth}{!}{
\small
\begin{tabular}{ccccc}
    \toprule \vspace{-1em}\\
    \textbf{Time}&\multicolumn{4}{c}{\textbf{Methods}}\\
   {$\begin{array}{c} \text {Percentile of} \\\text {follow-up time}\\\end{array}$} & G-formula & AIPTW &  Plug-in & One-step  \\
    \hline
    \rowcolor{gray!30} & \multicolumn{4}{c}{Weak unmeasured confounding effect} \\   
    30th & 93.4\% & 94.6\% & 97.8\% & 95.8\% \\ 
    40th & 94.8\% & 95.0\% & 97.4\% & 95.4\% \\ 
    50th  & 95.0\% & 93.2\% & 96.4\% & 94.6\% \\ 
    60th & 94.8\% & 91.8\% & 94.6\% & 92.8\% \\ 
   \hline \rowcolor{gray!30} & \multicolumn{4}{c}{Strong unmeasured confounding effect} \\
    30th & 0.0\% & 40.8\% & 72.0\% & 96.2\% \\
     40th & 0.2\% & 36.4\% & 76.8\% & 95.6\% \\ 
     50th  & 0.4\% & 34.4\% & 77.8\% & 95.0\% \\ 
     60th & 6.0\% & 35.8\% & 78.4\% & 95.4\% \\ 
    \bottomrule
\end{tabular}}
\end{center}
\RaggedRight {\footnotesize Note: Coverage probabilities were calculated based on $500$ simulated data sets with a sample size of $1000$. With strong unmeasured confounding effects, only the proposed one-step estimator maintains 95\% coverage.\\
\text{}}
\end{table}

Table \ref{c3tb:adjustUcov} reports the percentage of true ATE covered by the 95\% CIs across $500$ simulations at four follow-up time points. When unmeasured confounding effect is negligible, all methods consistently achieve around 95\% coverage. However, with strong unmeasured confounding effects, only the proposed one-step estimator maintains 95\% coverage, highlighting its ability to adjust for unmeasured confounding, and demonstrating its asymptotic normality for valid inference.

\subsection{Simulation set 2: Different model specifications}\label{c3ss:sim2}
After validating the ability to adjust for unmeasured confounding in time-to-event data, we further investigated the performance of the estimators when some of the nuisance functions were misspecified, a common challenge in real-world analyses where true models are often unknown. In this simulation, the binary instrumental variable $Z \sim Ber(\left[1+\exp \left\{-(0.2X_1-0.8X_2+0.2X_3+0.1X_4)\right\}\right]^{-1})$; the binary treatment $A$ was generated from $Ber(\left[1+0.3\exp\left\{-5(0.5X_1-2.8X_2+0.5X_3+0.1X_4)\right\}\right]^{-1}+0.63(Z-0.5)+0.07(U-0.5)+0.35)$; event and censoring time were generated from exponential distributions with means of $0.1\exp(0.1X_1+0.45X_2-0.15X_3+0.05X_4-1.5A-0.4U)$ and $0.5\exp(0.1X_1-0.5X_2+0.1X_3+0.05X_4+0.5A)$, respectively. 

In scenarios with model misspecification, certain covariates were deliberately omitted from fitting misspecified models for nuisances. Additionally, instead of using the true baseline covariates, a set of transformed baseline covariates were included in the misspecified models: $W_1=\exp(X_1/2), W_3=(X_1X_3/25+0.6)^3,\text{ and } W_4=(X_4+20)^2/50$. 

\begin{figure} 
\centerline{\includegraphics[width=0.8\textwidth]{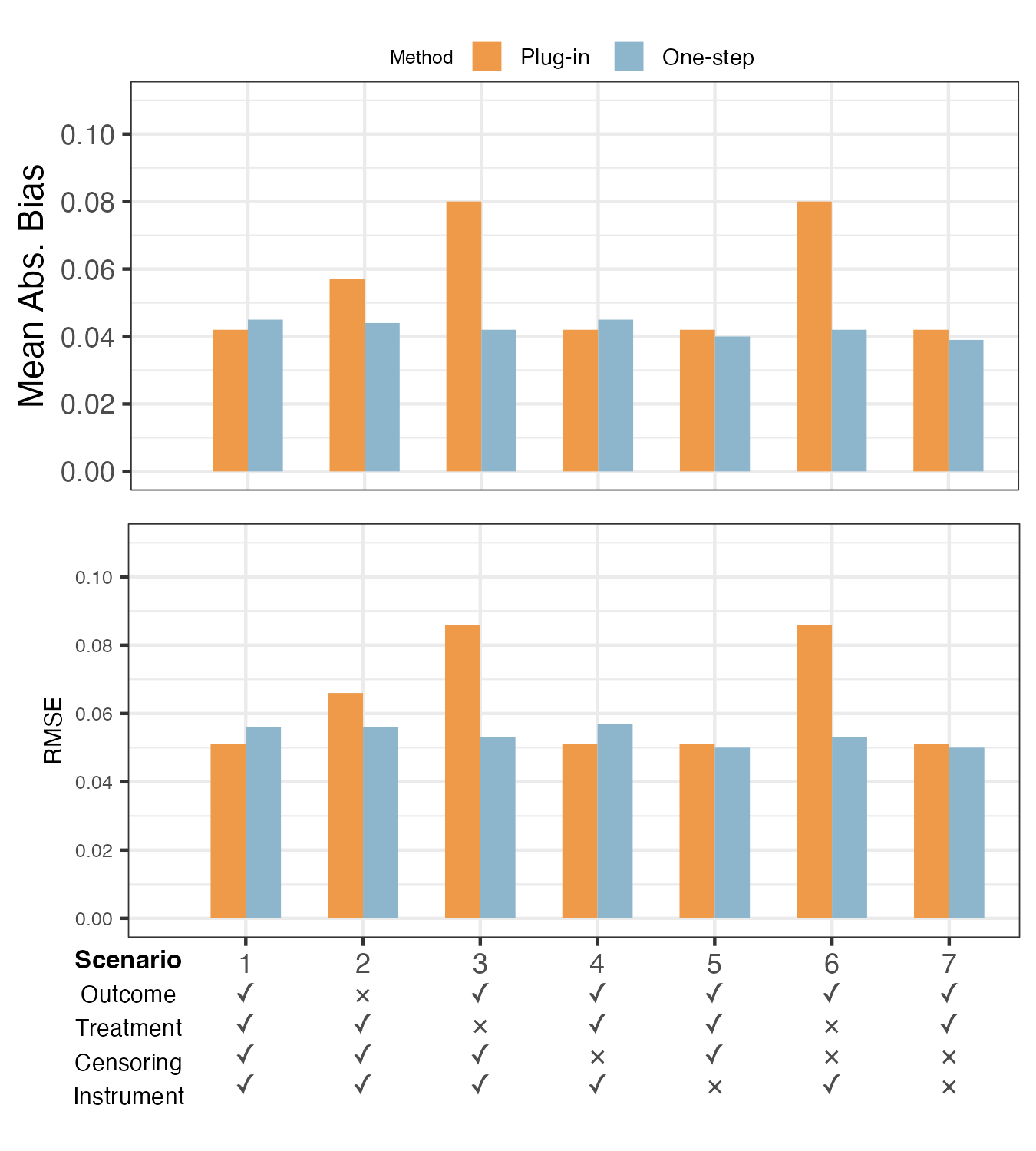}}
\caption[Performance of estimating ATE at the 50th percentile follow-up time under different model specification.]{Performance of estimating ATE at the 50th percentile follow-up time under different model specifications is summarized. The mean absolute bias and RMSE were calculated based on 500 simulated datasets, each with a sample size of 1000. Both the plug-in estimator and the proposed one-step estimator utilize the Cox model to estimate the CIF. For all estimators, censoring probability was estimated using a Cox model, while treatment probability and instrument probability were estimated using logistic regression when necessary. $\checkmark$ and $\times$ indicate whether the corresponding models were correctly specified and misspecified, respectively. Baseline covariates were deliberately omitted from fitting misspecified models: $X_2$ was omitted in the misspecified outcome model, $X_2$ and $X_4$ were omitted in the misspecified treatment model, $X_2$ was omitted in the misspecified censoring model, and $X_2$ was omitted in the misspecified instrumental variable model. A set of transformed baseline covariates were included in the misspecified models: $W_1=\exp(X_1/2), W_3=(X_1X_3/25+0.6)^3,\text{ and } W_4=(X_4+20)^2/50$. 
}
\label{c3f:DRBiasRMSE}
\end{figure} 

Figure \ref{c3f:DRBiasRMSE} shows that the proposed one-step estimator outperforms the plug-in estimator, especially when the outcome or treatment models are misspecified, with lower bias and RMSE. In Scenario 1, where all models are correctly specified, both perform well. However, in Scenarios 2 and 3, with misspecified outcome and treatment models, respectively, the plug-in estimator exhibits a significant increase in mean absolute bias and RMSE compared to Scenario 1. Specifically, when the treatment model is misspecified, the plug-in estimator's mean absolute bias increases by 100\%, and RMSE increased by approximately 40\%.  

Notably, in Scenarios 2 through 7, the proposed one-step estimator maintains the same level of mean absolute bias and RMSE as in Scenario 1, reflecting its double robustness, as described in  (\ref{c3eq:remainder}).  This indicates that as long as at least one nuisance component in each product term is consistent, the estimator remains consistent.


\subsection{Simulation set 3: Complex true models}\label{c3ss:sim3}
Real-world data are often too complex to be captured by parametric models. In this simulation set, we compare the performance of estimators incorporating parametric versus nonparametric models for data generated from complex true models. The binary instrumental variable $Z$ was generated from a logistic model reflecting this complexity:
 $Ber(\left[1+\exp \left\{-(0.4X_1X_2-0.8X_2^3+0.15X_3X_1+0.15sin(X_4)+0.4)\right\}\right]^{-1} )$, and the binary treatment $A\sim Ber(\left[1+0.15\exp\left\{-5(-X_1X_2-\right.\right.\left.\left.2X_2^3+0.75X_1X_3+0.15sin(X_4))\right\}\right]^{-1}+0.64(Z-0.5)+0.32(U-0.3)^2+0.24)$. Event and censoring time were generated from exponential distributions with means $\{0.2\exp(0.8X_1X_2+2X_2^3-1.5X_1X_3-0.3sin(X_4)+1.5A$\\$-sin(-6U)-1\}$ and $\{0.02\exp(-0.8X_1X_2-2.4X_2^3+0.6X_1X_3-0.3sin(X_4)+0.5A\}$. 
 
For the estimators using parametric models, the nuisances were estimated by the same models as in Simulation Sets 1 and 2. For the estimators using nonarametric models, the CIF and censoring probabilities were estimated by super learner implemented in the R package \texttt{survSuperLearner}, which includes Cox models, exponential regressions, log-logistic regressions, and survival random forests. Treatment probability and instrument prevalence were estimated by super learner (R package \texttt{SuperLearner}), which includes GLMs, GAMs, random forests, multivariate adaptive polynomial spline regressions, and marginal means.

\begin{figure} 
\centerline{\includegraphics[width=0.75\textwidth]{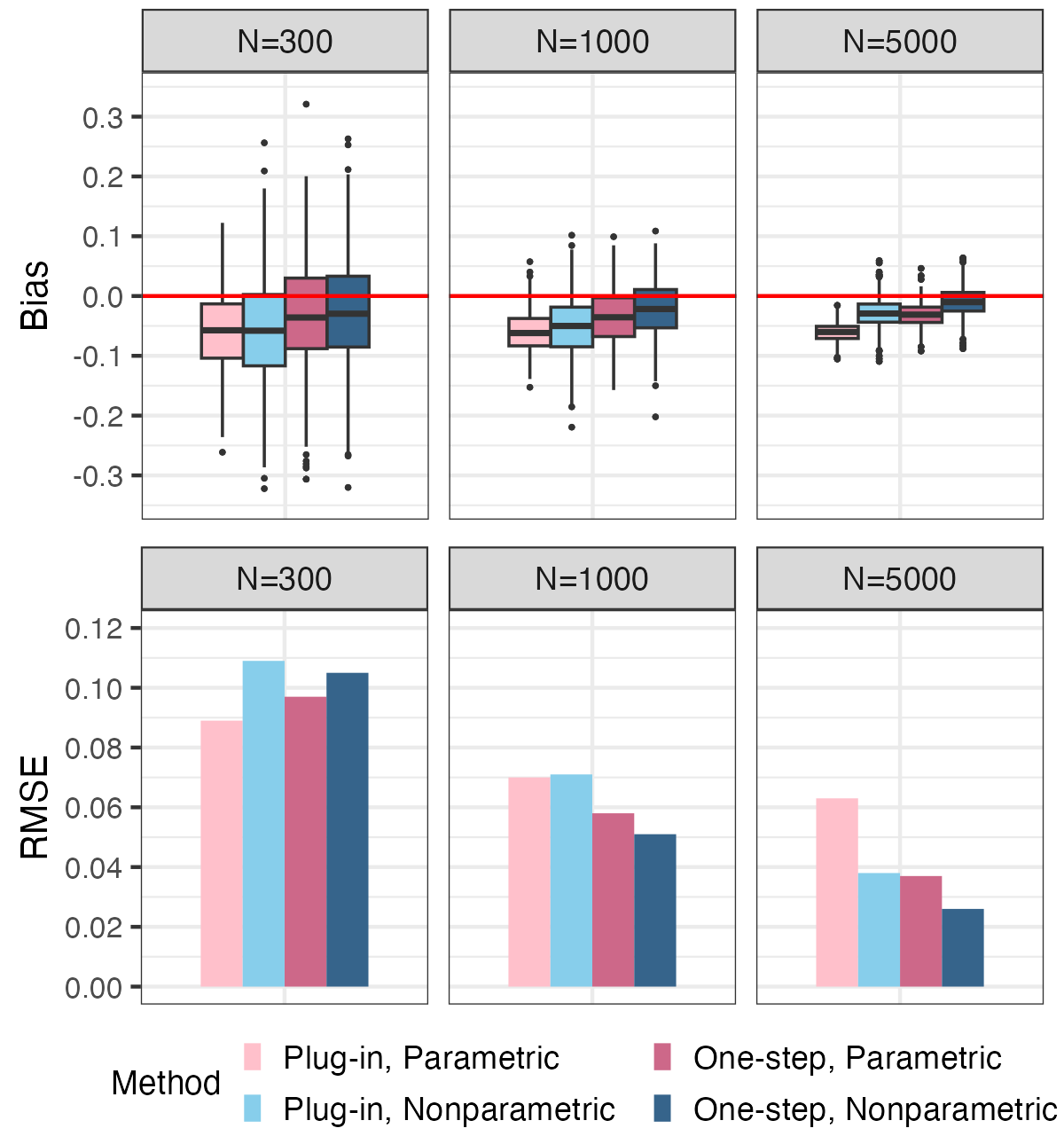}}
\caption[Performance of estimating ATE using parametric models and nonparametric models.]{Performance of estimating ATE at the 50th percentile follow-up time in data generated from complex true models. The bias and RMSE were calculated based on 500 simulated datasets,  with a sample size of 300, 1000 and 5000. For the plug-in estimator and the proposed one-step estimator incorporating parametric models, Cox models were used for estimating CIF and censoring probability, and logistic regressions were used for estimating treatment probability and instrument probability. For the estimators incorporating nonarametric models, the CIF and censoring probabilities were estimated using a super learner implemented in the R package \texttt{survSuperLearner}, which includes Cox models, exponential regressions, log-logistic regressions, and survival random forests. Treatment probability and instrumental variable prevalence were estimated using super learner implemented by the R package \texttt{SuperLearner}, which includes GLMs, GAMs, random forests, multivariate adaptive polynomial spline regressions, and marginal means.}
\label{c3f:NL} 
\end{figure} 

Figure \ref{c3f:NL} presents the bias and RMSE for the estimators using both parametric and nonparametric models, across different sample sizes. As the sample size increase, the estimators using nonparametric models exhibit smaller bias and RMSE than parametric ones, suggesting that nonparametric models can better capture the complex relationships. When sample size increase to 5000, only the proposed one-step estimator using non-parametric models provides unbiased estimates. This also highlights the flexibility of the proposed one-step estimator in accommodating nonparametric models, increasing accuracy for complex real-world data.

\section{Application to ICU Patients with Vasopressor-dependent Septic Shock}\label{c3s:app}
\subsection{Data Description and Motivation}
Septic shock is a severe, high-mortality condition requiring vasopressor support. Hydrocortisone is frequently used in vasopressor-dependent septic shock, but its impact on mortality remains uncertain due to conflicting study results \citep{li2024hydrocortisone,sprung2008hydrocortisone,zhang2023timing, venkatesh2019hydrocortisone}, likely influenced by unmeasured confounders. To address this, we applied our proposed IV method to assess the ATE of hydrocortisone on mortality.

Data were obtained from electronic medical records of patients admitted to the intensive care units (ICUs) of a healthcare system from 2018 to 2020, including 2,023 patients with vasopressor-dependent septic shock. Day 0 was defined as the day of vasopressor initiation, with follow-up censored at 30 days to focus on short-term treatment effects. Eligible patients were aged 18 or older, not designated as "comfort measures only”, survived beyond day 1, and had a maximum sequential organ failure assessment (SOFA) score of 7 or below.

The treatment variable was defined as whether hydrocortisone was administered on the day of or the day after vasopressor initiation. Covariates collected included patient demographics and clinical characteristics such as age, gender, SOFA score, mechanical ventilation use, maximum vasopressor dose, ICU admission source, and Elixhauser comorbidity index; attending physician characteristics (experience, sex, specialty); and ICU-level factors. 

Given the potential for unmeasured confounding, we used the attending physician’s preference for prescribing hydrocortisone as an instrumental variable \citep{rassen2009instrumental}. Physician preferences, measured as the proportion of septic shock patients treated with hydrocortisone, was categorized into high (above 18\%) and low (18\% or below) groups. To ensure IV validity, we excluded 73 patients whose attending physicians treated $\leq 5$ patients during the study period, resulting in an analysis sample of 1,882 patients. Among them, 306 patients (16.3\%) died within 30 days, and 333 patients (17.7\%) received hydrocortisone. Of the treated patients, 73\% were managed by physicians with high hydrocortisone prescribing preferences, compared to 47.3\% of the untreated patients. This strong association between physician preferences and treatment assignment supports the validity of our instrument.

\subsection{Data Analysis Results}
\begin{figure} 
\centerline{\includegraphics[width=\textwidth]{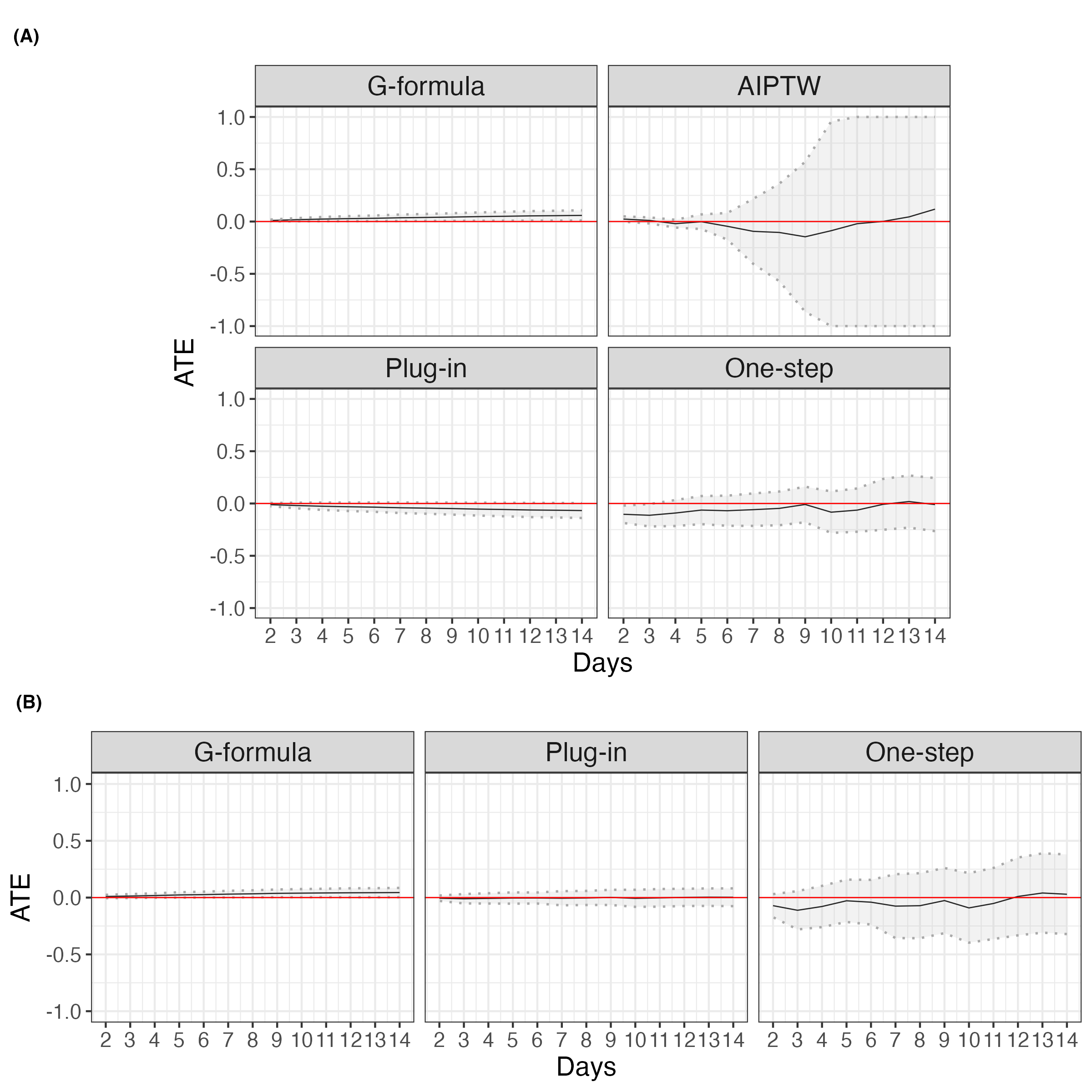}}
\caption{Average treatment effect (ATE) of hydrocortisone on patients with vasopressor-dependent septic shock from day 2 to day 14, estimated using the G-formula, the AIPTW estimator, the plug-in estimator, and the proposed one-step estimator. The ATE is defined as the difference in the cumulative incidence function (CIF) between treated and untreated patients. Shaded areas represent 95\% confidence intervals, calculated using the efficient influence function for the AIPTW and one-step estimators, and using the sampling standard error for the G-formula and plug-in estimators. The red line indicates no treatment effect.
Panel (A): All estimators use a Cox model for CIF estimation. The censoring probability was estimated using a Cox model, while treatment and instrument probabilities were estimated using logistic regression, where applicable. Panel (B): All estimators use super learner for CIF estimation. The censoring, treatment, and instrument probabilities were also estimated using super learner, where applicable.}
\label{c3f:ATE_para}
\end{figure}

To estimate the ATE of hydrocortisone on mortality, we applied our proposed one-step estimator and compared with the plug-in estimator, the G-formula, and the AIPTW estimator. The results are summarized in Figure \ref{c3f:ATE_para}, with parametric models in panel (A) and super learner for nuisance parameters in panel (B). 

In the parametric setting, the G-formula produces positive ATE estimates with 95\% CIs above zero, suggesting hydrocortisone increases mortality. However, our one-step estimator yields negative ATE estimates at most time points, with 95\% CIs including zero, indicating no significant effect. This discrepancy highlights the importance of addressing unmeasured confounding, which the G-formula does not account for. 

Both AIPTW and plug-in estimators generate negative ATE estimates, with 95\% CI containing zero. However, the AIPTW estimator shows a sharp increase in variance after day 6, making its estimates unreliable in smaller samples due to the imbalance in treatment and censoring groups. In contrast, our one-step estimator maintains stable performance, providing more reliable inference even with treatment imbalance.

When applying super learner for the nuisance functions, see panel (B), variances are slightly larger compared to parametric models, as expected, but the conclusions remain consistent. The ATE estimates from the G-formula are lower than in the parametric models, with zero in the 95\% CIs, suggesting no significant treatment effect. 

Overall, our findings suggest the presence of unmeasured confounding that must be to be accounted for when investigating the treatment effect of hydrocortisone on mortality. Based on the results from our proposed doubly robust estimator, we conclude that hydrocortisone does not significantly impact mortality in this patient population.

\section{Discussion}\label{c3s:dis}
We developed a novel instrumental variable (IV) method for estimating the causal ATE in time-to-event data with unmeasured confounders. The ATE, defined as the difference in potential cumulative incidence functions, provides an interpretable alternative to hazard ratios, avoiding the selection bias inherent in the latter. Our method extends to settings with competing risks, making it highly relevant for clinical data with multiple event types. 

Using semiparametric theory, we built a one-step estimator based on the efficient influence function, ensuring double robustness and asymptotic properties. This estimator remains unbiased even with misspecified nuisances, and supports valid inference based on its asymptotic normality, enabling straightforward confidence interval construction. Its flexibility surpasses existing IV methods, effectively addressing the complexity of real-world data.


A key strength of our approach is its ability to incorporate various models for nuisance components without relying on strict parametric assumptions. This enables the use of advanced machine learning models, making it ideal for complex real-world data. Our estimator also ensures reliable confidence intervals and faster convergence. Future work will focus on integrating deep learning methods to enhance its capacity for high-dimensional data, further advancing its utility in precision medicine.

\bibliographystyle{biom}
\bibliography{ref}

\newpage

\end{document}